\documentclass[atoms,article,submit,moreauthors,10pt,a4paper]{mdpi}
\usepackage{verbatim}
\firstpage{1} 
\articlenumber{x}
\doinum{10.3390/------}
\pubvolume{xx}
\pubyear{2016}
\copyrightyear{2016}
\externaleditor{Academic Editor:  Elmar Tr\"abert }
\history{Received: date; Accepted: date; Published: date}

\Title{Electroweak Decay Studies of Highly Charged Radioactive Ions with TITAN at TRIUMF}

\Author{K.G.~Leach $^{1}$*, I.~Dillmann$^{2}$, R.~Klawitter$^{2,3}$, E.~Leistenschneider$^{2,4}$, A.~Lennarz$^{2}$, T.~Brunner$^{2,5}$, D.~Frekers$^{6}$, C.~Andreiou$^{7}$, A.A.~Kwiatkowski$^{2}$, and J.~Dilling$^{2}$}

\address{%
$^{1}$ \quad Department of Physics, Colorado School of Mines, Golden, CO, 80401 USA\\
$^{2}$ \quad TRIUMF, Vancouver, BC, V6T 2A3 Canada\\
$^{3}$ \quad Max-Planck-Institut f\"ur Kernphysik, Saupfercheckweg, D-69117 Heidelberg, Germany\\
$^{4}$ \quad Department of Physics and Astronomy, University of British Columbia, Vancouver, BC, V6T 1Z1 Canada\\
$^{5}$ \quad Department of Physics, McGill University, Montr\'eal, QC, H3A 2T8 Canada\\
$^{6}$ \quad Institut f\"ur Kernphysik, Westfalische Wilhelms-Universit\"at, D-48149 M\"unster, Germany\\
$^{7}$ \quad Department of Chemistry, Simon Fraser University, Burnaby, BC, V5A 1S6 Canada
}

\corres{Correspondence: kleach@mines.edu}

\abstract{Several modes of electroweak radioactive decay require an interaction between the nucleus and bound electrons within the constituent atom.  Thus, the probabilities of the respective decays are not only influenced by the structure of the initial and final states in the nucleus, but can also depend strongly on the atomic charge.  Conditions suitable for the partial or complete ionization of these rare isotopes occur naturally in hot, dense astrophysical environments, but can also be artificially generated in the laboratory to selectively block certain radioactive decay modes.  Direct experimental studies on such scenarios are extremely difficult due to the laboratory conditions required to generate and store radioactive ions at high charge states.  A new electron-beam ion trap (EBIT) decay setup with the TITAN experiment at TRIUMF has successfully demonstrated such techniques for performing spectroscopy on the radioactive decay of highly charged ions.}


\begin{document}


\section{Introduction}
Most forms of nuclear decay involve only the bound protons and neutrons that constitute the atomic nucleus, and require little-to-no interactions with the electrons that typically surround them.  However, some common modes of electroweak decay such as orbital electron capture (EC) and internal electron conversion (IC), proceed through an interaction between the nucleus and bound electrons within the constituent atom~\cite{KRANE}.  Additionally, for radioactive decay modes that emit charged leptons ($\beta^+$/$\beta^-$ decay which emit positrons/electrons, respectively), interactions with the surrounding electron cloud of the atom can change the energy and shape of the observed particle-emission momentum distributions.  As a result, these respective decay modes are not only influenced by the structure of the initial and final states in the nucleus, but can also depend strongly on the atomic charge state~\cite{Lit11}.  These effects, particularly for EC and IC, become increasingly more significant as the atom is ionized closer to electron shells with the largest spatial overlap with the nucleus ($K$ and $L$ shell).

In general, very few experimental studies have been performed on the decay of highly charged ions (HCIs), largely due to significant technical obstacles of creating and storing radioactive nuclides at high atomic charge states.  In fact, the vast majority of measurements in this sub-field of low-energy nuclear physics have been performed by one group using the Experimental Storage Ring (ESR) at GSI, in Darmstadt, Germany.  Their work over the last few decades has been primarily focused on the effects of high charge states (bare, H-like, He-like, etc) on electron capture using Schottky mass spectrometry~\cite{Lit11}.

From a fundamental physics standpoint, these measurements are motivated by:
\begin{enumerate}
\item Tests of the electroweak interaction under varying atomic conditions, and
\vspace{5pt}
\item Investigating nucleosynthesis under extreme astrophysical environments which are hot enough to partially or fully strip atoms of their bound electrons.
\end{enumerate}

For several years there has been increased interest in trying to perform similar measurements of HCIs in ion traps at low energies due to the increased experimental control over the decay environment.  This article reports the progress of such work using the only low-energy ion trap in the world capable of performing experiments on the decay of HCIs.  The sections below highlight the trap itself, and outline a sample of the research directions which are planned in the near future.

\section{The TITAN Facility at TRIUMF-ISAC}
\begin{figure}[t!]
\centering
\includegraphics[width=0.5\linewidth]{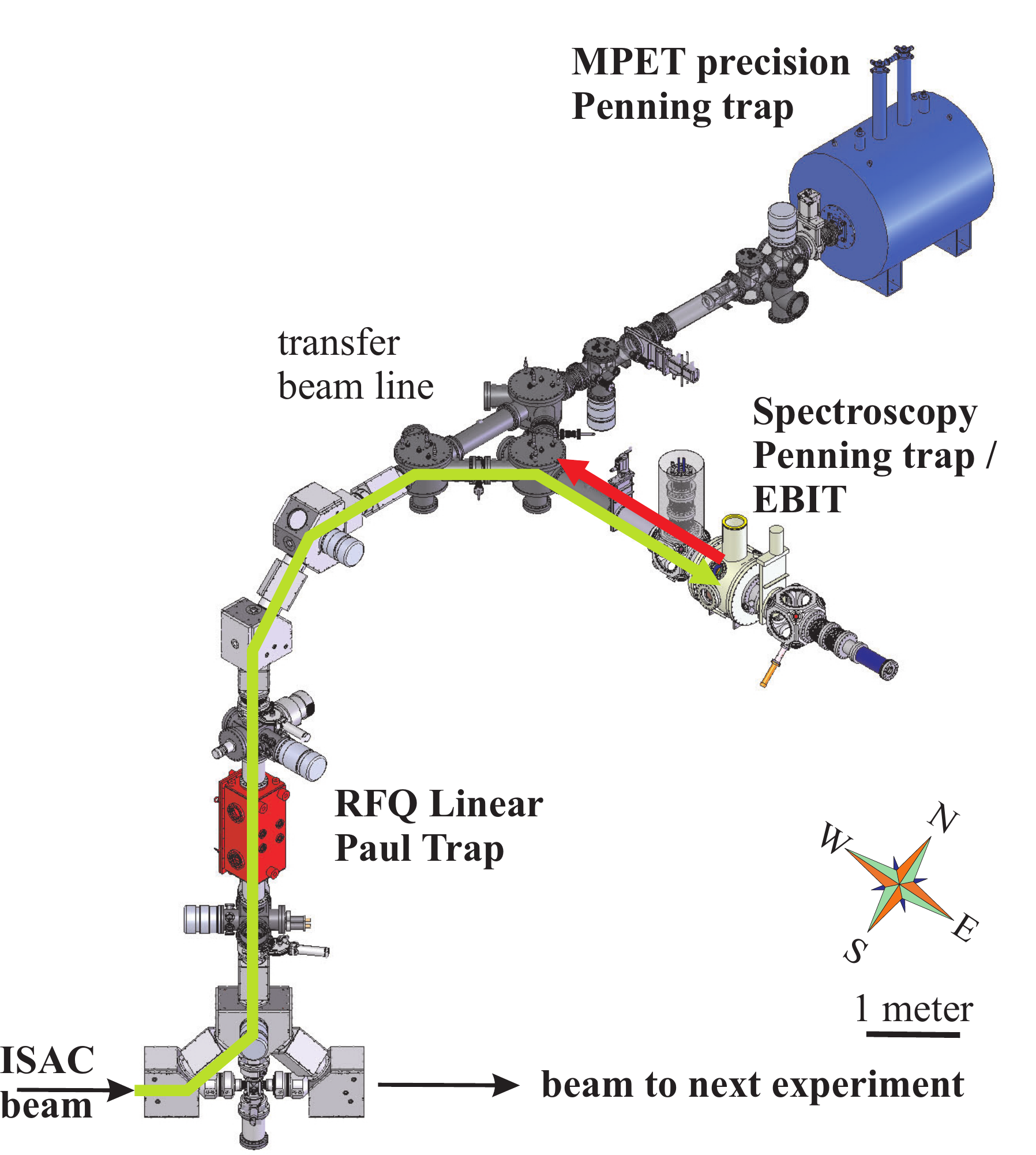}\includegraphics[width=0.45\linewidth]{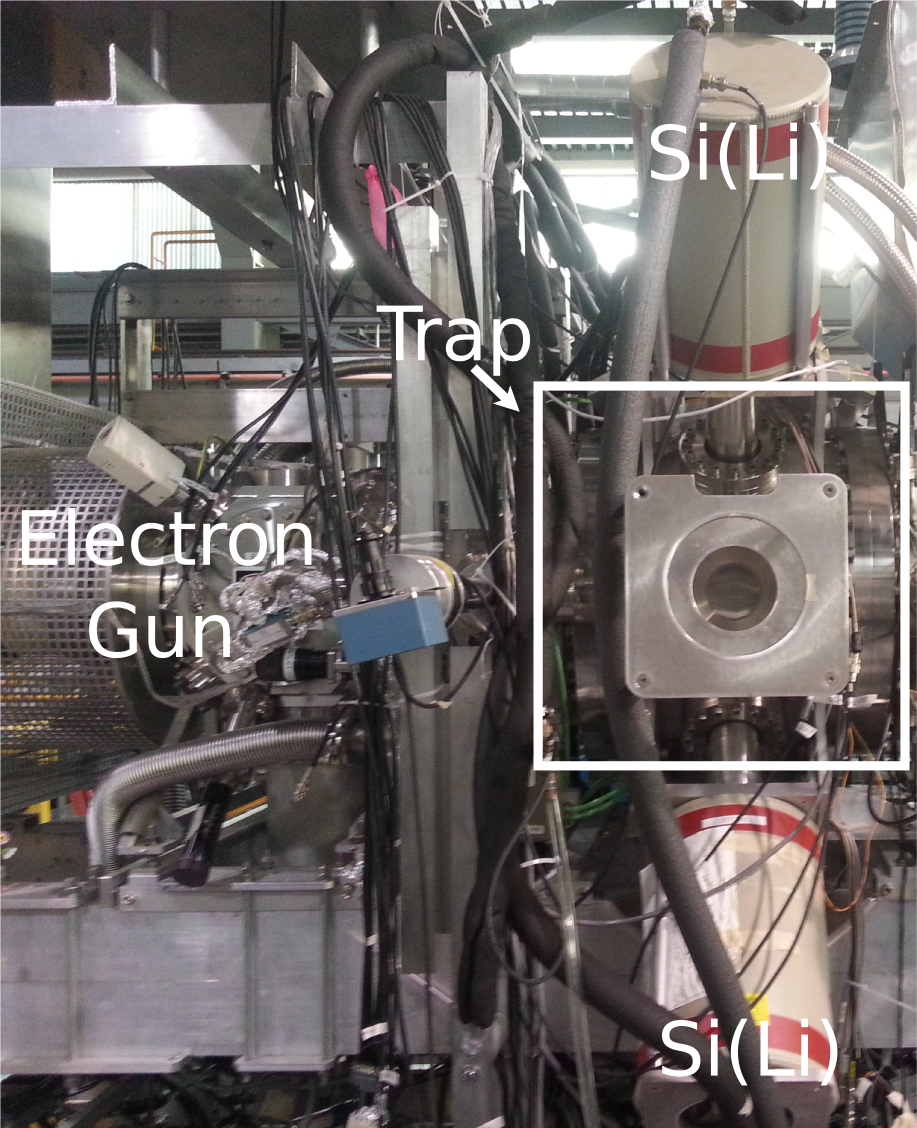}
\caption{\label{TITAN}(left) A schematic of the current TITAN experimental setup in the ISAC facility at TRIUMF.  The injection of bunched, singly charged ions into the EBIT is displayed by the green line, while the extraction of HCIs is shown by the red line. (right) A photograph of the EBIT decay spectroscopy setup in its configuration on the TITAN platform.  Shown are the electron gun, two of the seven Si(Li) detectors, and the trap outline.}
\end{figure}
Due to the typically short decay half-lives of the radioactive nuclides of interest ($T_{1/2}\approx10^{-3}-10^{3}$~s), they must be produced, purified, and delivered to the respective experimental setups in a short amount of time.  The Isotope Separator and Accelerator (ISAC) facility~\cite{Bri97} at TRIUMF in Vancouver, Canada, employs a high-intensity (up to 100 $\mu$A) beam of 500~MeV protons to produce radioactive ion beams (RIBs) using the isotope separation on-line (ISOL) technique~\cite{Blu13}.  ISAC is currently able to provide a wide variety of RIBs via spallation and fission reactions through the use of several different production target and ion-source combinations~\cite{Dil14}.  Following the in-target production the ions are mass separated for isobaric purity before being delivered to the experimental facilities for research using RIBs.  For the decay studies described in this article, the mass-selected continuous beam of radioactive singly charged ions (SCIs) is delivered at low energies ($<60$~keV) to TRIUMF's Ion Trap for Atomic and Nuclear Science (TITAN)~\cite{Dil06}.  The TITAN system consists of three ion traps:
\begin{enumerate} 
\item A radio-frequency quadrupole (RFQ) linear Paul trap~\cite{Smi06} for buffer-gas cooling and bunching the low-energy ion beam,
\item A 3.7~T, high-precision mass-measurement Penning trap (MPET)~\cite{Bro12}, and
\item An electron-beam ion trap (EBIT) which is used to create HCIs~\cite{Lap10}, and for performing decay spectroscopy on trapped radioactive nuclei~\cite{Lea15}.
\end{enumerate}

Two additional ion traps will be added to the TITAN system in the near future; a multi-reflection time-of-flight (MR-ToF) isobar separator~\cite{Jes15}, and a cooler Penning trap (CPET) for cooling the HCIs before injection into MPET for higher precision atomic mass measurements~\cite{Sim11}.  A schematic view of the TITAN facility at TRIUMF-ISAC is shown in Fig.~\ref{TITAN}, along with the ion path for typical decay-spectroscopy experiments.

\subsection{The TITAN EBIT}
\begin{figure}[t!]
\centering
\includegraphics[width=0.5\linewidth]{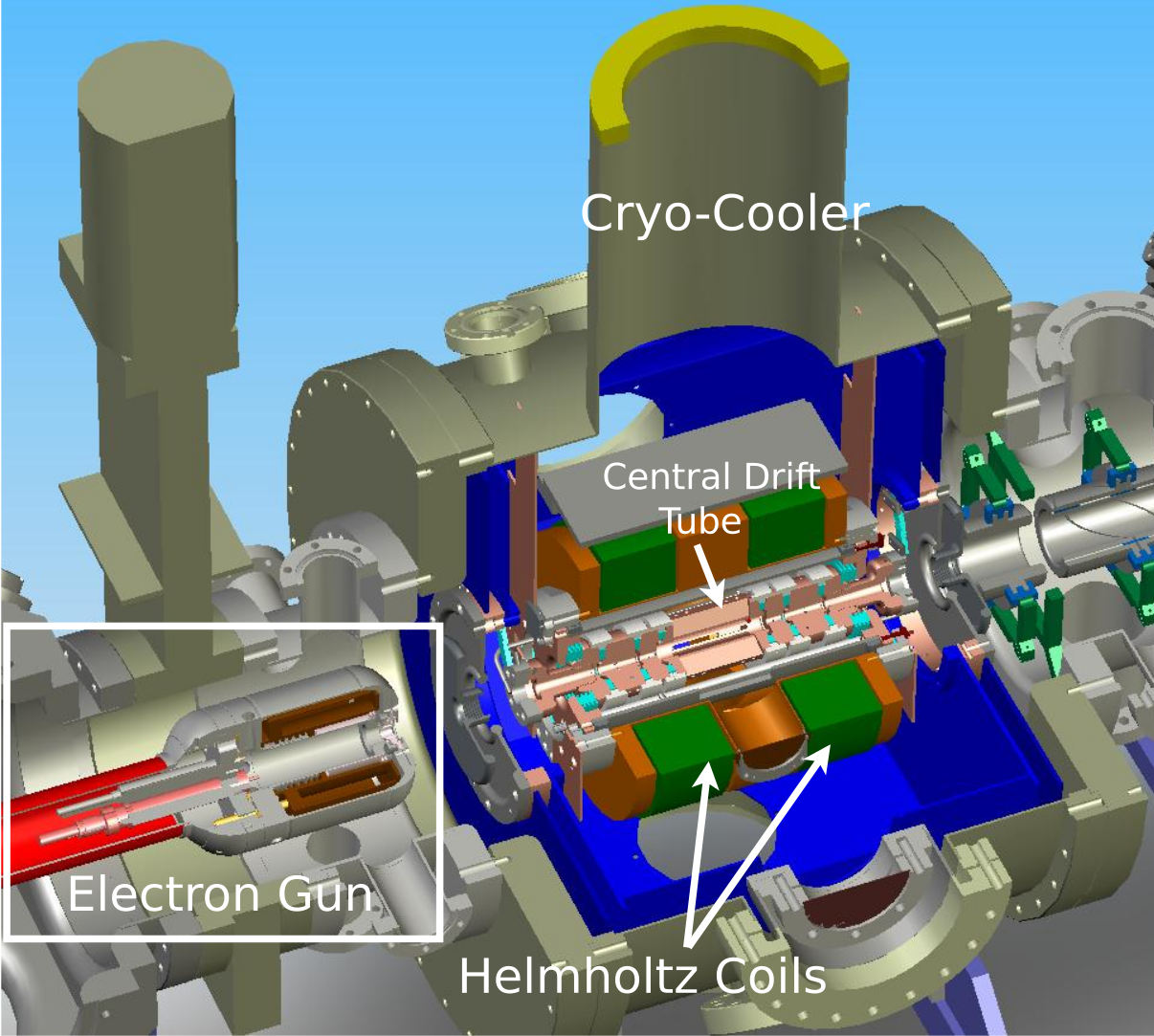}~~~~~~~\includegraphics[width=0.5\linewidth]{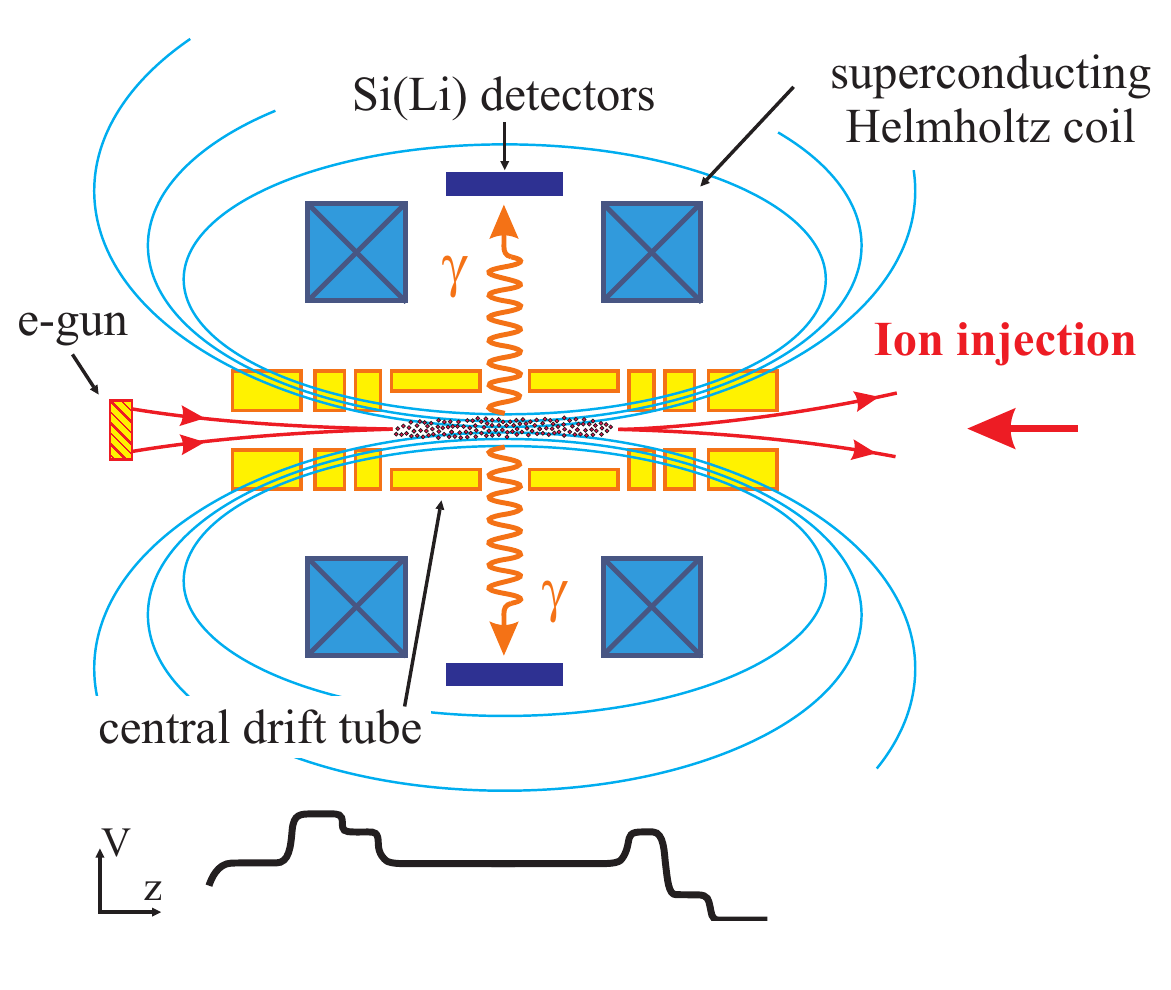}
\caption{\label{trap_schematic}(left) A schematic view of the trap and electron-gun portions of the TITAN EBIT.  (right) An illustration of the EBIT operation for decay spectroscopy and charge breeding (not to scale).  Depicted below the trap components is a typical potential profile of the trap electrodes while the ion bunch is confined in the EBIT.}
\end{figure}
The TITAN EBIT, shown in Fig.~\ref{trap_schematic} generates highly charged ions through the process of electron-impact ionization~\cite{EI}.  The trap itself consists of an electron gun, a cold drift-tube assembly which is thermally coupled to a superconducting magnet, and an electron collector.  This concept follows the successful FLASH-EBIT design from the Max-Planck-Institut f\"ur Kernphysik in Heidelberg, Germany, where the system is used for experiments on highly charged stable ions~\cite{Epp07}.  The trapped ions are axially confined by an electrostatic square-well potential formed by applying voltage to the drift tubes.  The typical voltage profile employed for trapping and breeding the ions is shown in the lower part of Fig.~\ref{trap_schematic} (right).  Radial confinement is provided by both the electron-beam space-charge potential, and an up-to 6~Tesla magnetic field produced by two superconducting coils in a Helmholtz-like configuration~\cite{Lap10}.  The total ion trapping capacity of the EBIT is determined by the electron beam negative space charge.  The number of negative elementary charges within the central trap region depends linearly on the electron beam current, and inversely on the square root of its energy.  For the TITAN EBIT, with an up-to 500 mA, 2~keV electron beam, the trapping region contains roughly $10^9$ electrons.  Thus, for an average ion charge state of $q\approx30^+$ in the nuclei of interest for this program, roughly $10^7$ ions can be confined in the trap.  This has been demonstrated and confirmed previously during the commissioning experiments.  Under these conditions, ion losses are very small (none have been observed previously), and the cycling of the charge state of the ion due to successive ionization and recombination processes does not affect the ion inventory.

\subsubsection{Decay Spectroscopy of HCIs with TITAN}
\begin{figure}[t!]
\centering
\includegraphics[width=0.42\linewidth]{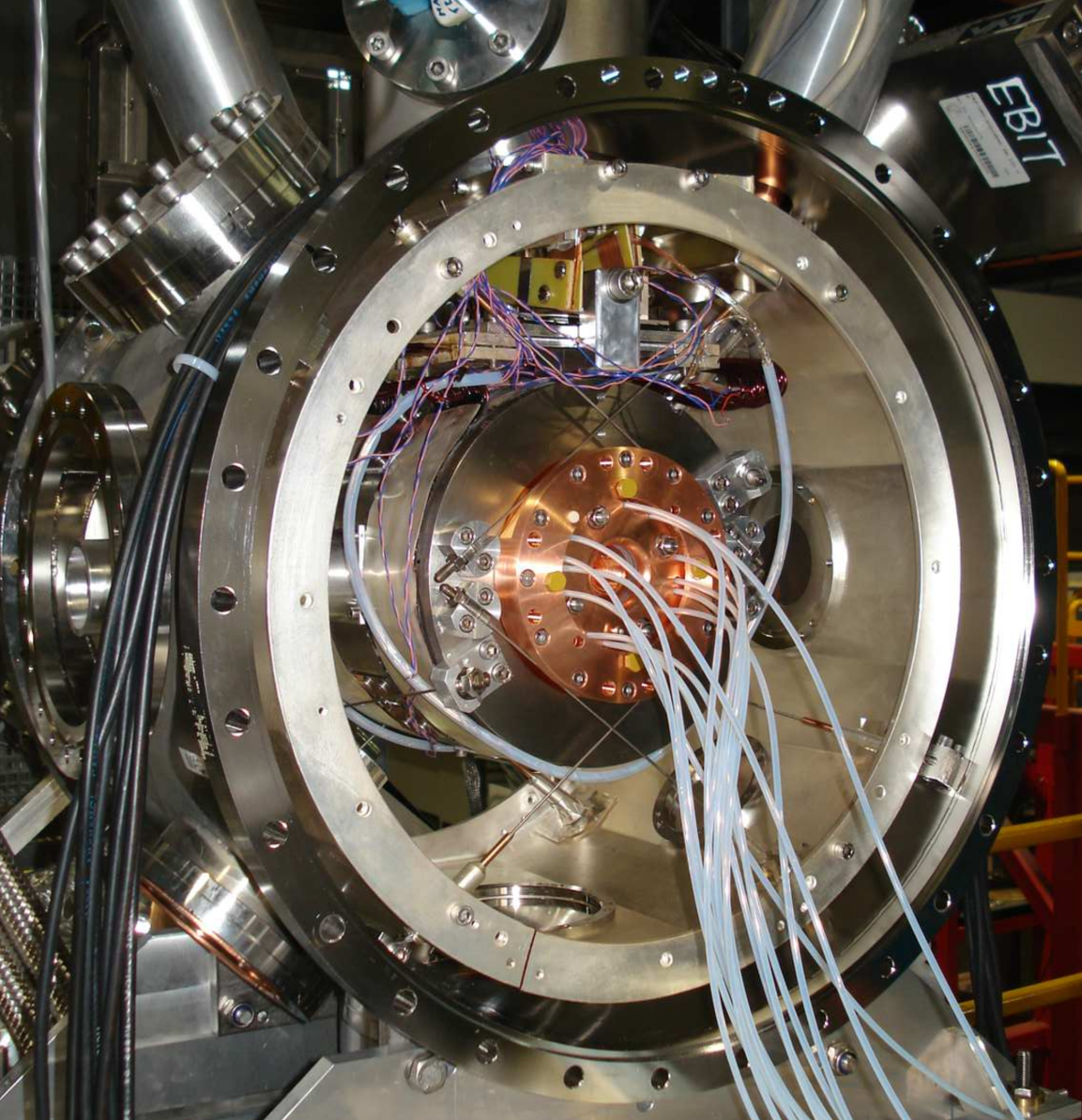}~~~\includegraphics[width=0.55\linewidth]{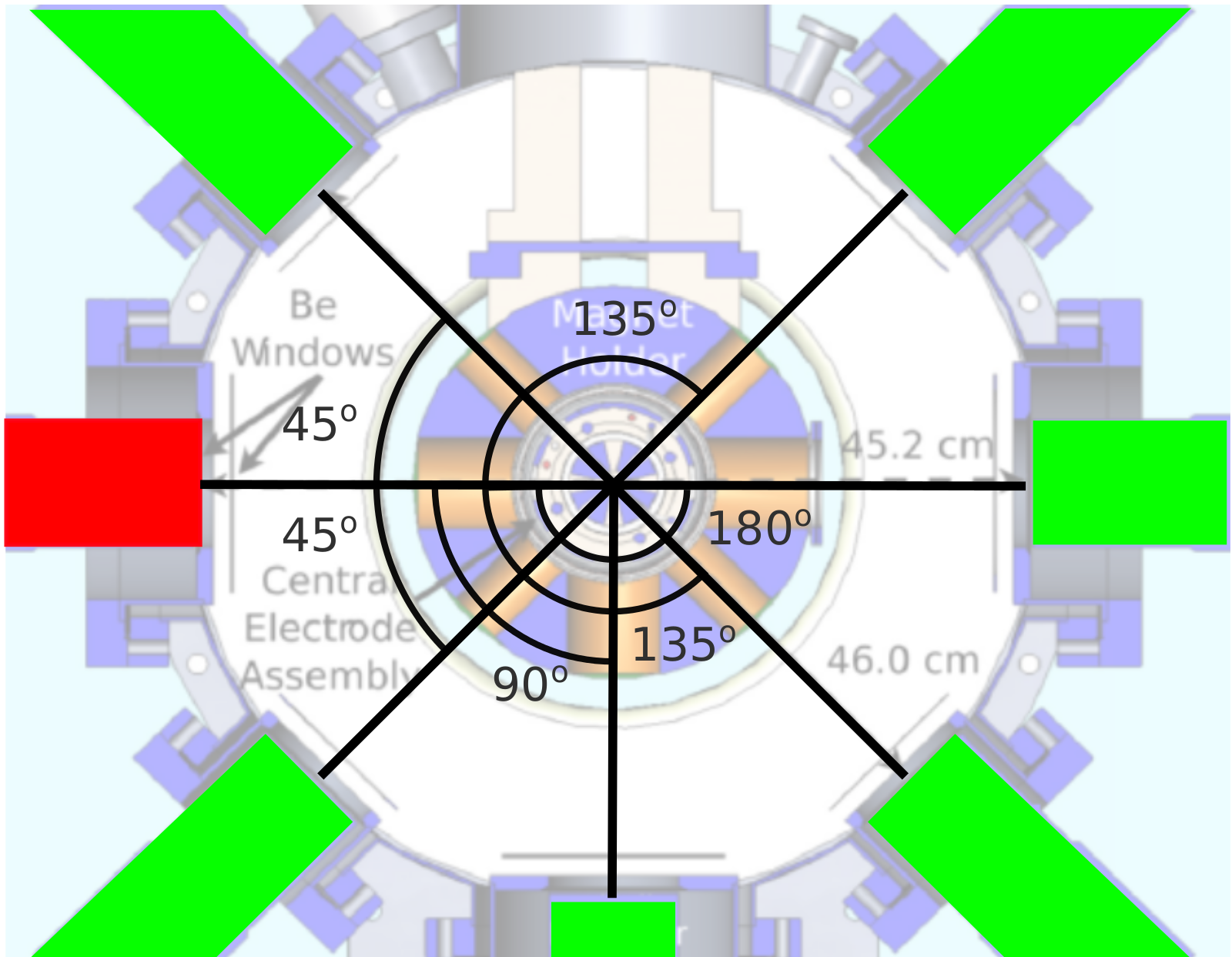}
\caption{(left) An image of the interior of the TITAN EBIT with the end-cap off as seen from the beam-injection side.  (right) A cross-sectional schematic of the TITAN EBIT showing the seven access ports and their relative angles to the horizontal axis, as depicted by the red detector.}
\label{EBIT_angles}
\end{figure}
The EBIT features seven external ports separated by $45^\circ$ from each other (Fig.~\ref{EBIT_angles} (right)), each with a 35~mm radius opening that is separated from the vacuum of the trap by 0.25~mm thick beryllium windows.  For performing decay-spectroscopy experiments on HCIs, these ports can house individual photon detectors that allow for high-sensitivity characterization of radioactive decay~\cite{Lea15}.  Additionally, the in-trap composition of ion species and charge states can be periodically monitored by pulsing the trap contents out to a micro-channel plate (MCP) detector that is located downstream of the EBIT in the TITAN beamline.  The EBIT is operated in a cycled mode which typically consists of three parts: injection, storage/trapping, and extraction.  The cycles are optimized to increase the photon signal-to-background ratio from the atomic transitions in the species of interest, and thus trapping portions of the cycle can last anywhere from a few seconds to minutes.  Long trapping times are particularly important here, as the concept for EBIT-based decay spectroscopy with TITAN was first envisaged in the context of nuclear structure measurements for neutrinoless double beta decay~\cite{Fre07}.

Two commissioning measurements with the current setup have already been performed: $^{124}$Cs EC decay and $^{116}$In IC decay, which are reported in Refs.~\cite{Len14} and \cite{Lea15b}, respectively.  For these experiments, charge states of $q=28^+$ and $q=22^+$ were used, corresponding to stripping electrons near the respective atomic $N$ shells.  Despite the relatively large number of electrons that remained, atomic effects on the order of $\sim100$~eV were observed, and demonstrated the high-level of sensitivity achievable with this setup~\cite{Len14,Lea15b}.  


\section{Study of HCIs in Astrophysical Scenarios}
\begin{figure}[t!]
  \centering
  \includegraphics[width=\linewidth]{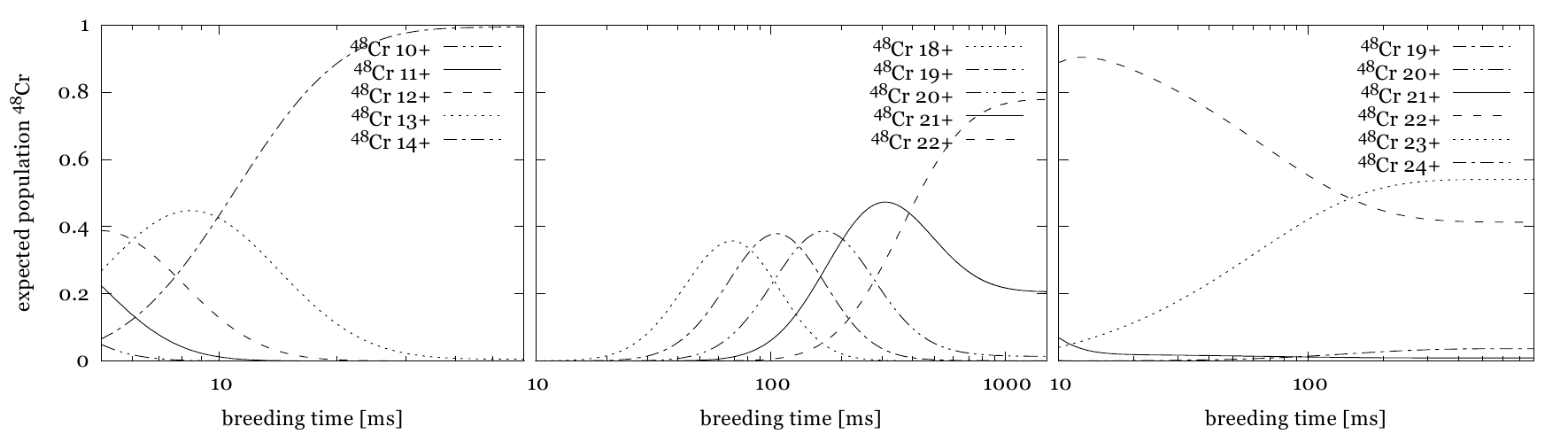}\\
  \includegraphics[width=\linewidth]{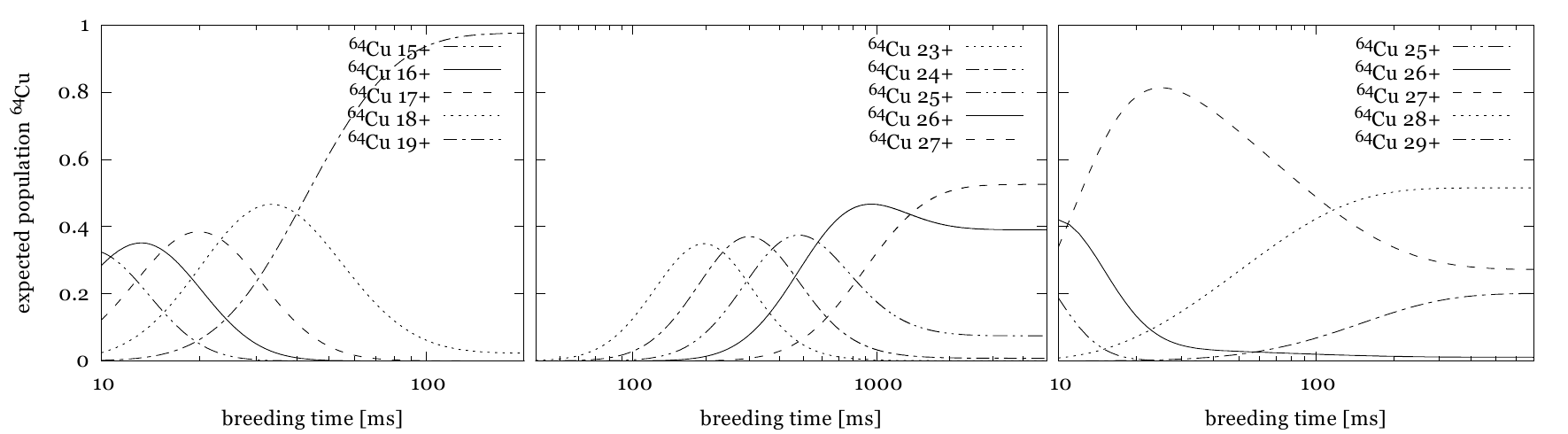}
  \caption{\small Simulated charge-state distributions using EBIT parameters optimized for the (from left to right) Ne-like, He-like and H-like charge states of $^{48}\mathrm{Cr}$ (top row) and $^{64}\mathrm{Cu}$ (bottom row).  The H-like distributions are simulated with parameters of a reconfigured TITAN-EBIT, namely $2~\mathrm{A}$ at $14~\mathrm{keV}$ and a 50~$\mu$m electron-beam diameter.}
  \label{ch_dists}
\end{figure}
The hindrance or full blocking of the EC decay mode can occur in several astrophysical environments that are hot enough to fully (or partially) ionize the radioactive atoms.  Perhaps the most prominent case is that of $^{7}$Be in the core of the sun, where temperatures can exceed $1.5\times10^7$~K ($\sim1.3$~keV) which is hot enough to fully ionize beryllium.  Since the $^{7}$Be decay is an integral step of the {\it pp-II} chain for hydrogen burning in the sun, the alteration of its half-life through ionization is a particularly relevant topic for determining how stars burn~\cite{Bah62,Gru97}.  Since the final state in EC decay has only two bodies (recoil nucleus and neutrino), the rate at which these nuclei decay is also particularly important for understanding the solar neutrino spectrum.  The terrestrial half-life for $^{7}$Be is $T_{1/2}=53.3$~days.  However, since the only mode of radioactive decay is through electron capture, a fully ionized $^{7}$Be$^{4+}$ can effectively become stable\footnote{In practice, the electron densities in the sun are high enough that $^{7}$Be$^{4+}$ can capture a free electron from the surrounding plasma and still undergo radioactive decay, as evidenced by the observation of the two lines in the solar neutrino spectrum at 861~keV and 384~keV~\cite{Ade11,Sim13}.}.

For higher-energy scenarios in the cosmos, such as the acceleration of material by supernovae (SN) explosions to relativistic energies ($>$200~MeV/$u$) of galactic cosmic-ray particles, even heavy atomic systems with large numbers of bound electrons can be fully ionized.  Through this method, relatively short-lived isotopes such as $^{37}$Ar (t$_{1/2}$= 35~days) and $^{51}$Cr (t$_{1/2}$= 27.7~days) which are produced by cosmic-ray spallation reactions on the SN ejecta can become long-lived ($\approx10^6$~years) due to the suppression of their EC decay mode.  As the lifetimes of these nuclides become increasingly longer, they can provide more complete information on distant astrophysical phenomena as they are detected by earth-bound satellites collecting cosmic-ray particles~\cite{Sto98}.  If a weak decay branch exists in these cosmic isotopes besides the dominant EC-branch, the removal of all electrons leads to a long partial stellar half-life governed only by the decay of the weak branch.  Prominent examples are $^{54}$Mn (EC, $\beta^+$/$\beta^-$), $^{56}$Ni (EC, $\beta^+$), and $^{59}$Ni (EC, $\beta^+$).  Several measurements of these weak decay branches have been performed under terrestrial conditions, mainly via $\beta^\pm$- or $\gamma$-spectroscopy, see e.g.~\cite{Zae99}, but very few (or none) have been performed under these proper astrophysical conditions.  Under those conditions, the EC lifetimes do not become significantly affected until the atom is nearly fully stripped of its bound electrons.  In fact, in most cases, orbital EC is only strongly influenced by the density of electrons in the $K$ and $L$ atomic shells. A simplified view can be described as follows:
\begin{itemize}
\item	No orbital electrons (bare nucleus): the isotope is stable unless free electrons are captured or other decay modes are possible (see $^7$Be$^{4+}$ in the sun).
\item	One $s$-electron (H-like state): a single electron in the $K$ shell can be captured.  Since only half of the $K$ shell is occupied, the decay half-life should be twice as long.
\item	Two orbital $s$-electrons (He-like state): both electrons are in the $K$ shell.  The decay rate roughly corresponds to the terrestrial rate, but without contributions from $L$-shell capture (proportional to $n^{-3}$).
\end{itemize}
However, this picture is not universally true. It was shown in storage ring measurements at GSI Darmstadt that for H- and He-like $^{122}$I$^{52+,51+}$~\cite{Ata12}, $^{140}$Pr$^{58+,57+}$, and $^{142}$Pm$^{60+,59+}$ ions~\cite{Lit07,Win09} that these simple assumptions do not hold for HCIs.  In all three cases, the He-like ions decay with the predicted half-life of $\frac{9}{8}$$\cdot$t$_{1/2}$(terr.), while the behaviour of the H-like ions is strongly dependent on the hyperfine structures of the initial and final nucleus. For $^{140}$Pr$^{58+}$ and $^{142}$Pm$^{60+}$, the measured half-life became even shorter than the terrestrial half-life: $\frac{9}{10}$$\cdot$t$_{1/2}$(terr.), and not twice as long as in the simplified picture.  An explanation for the observed deviation was presented in Ref.~\cite{Pat08} for allowed transitions, and states that the conservation of total angular momentum in the nucleus-lepton system has to be taken into account.  This is necessary since only certain spin orientations of the nucleus and of the captured electron can contribute to the allowed decay. The suppression or enhancement for a single-electron ion depends on the populated hyperfine state.  In all three of the previously investigate cases, the dominant EC branch were $1^+\rightarrow0^+$ ground-state to ground-state transitions.  As a result of these interesting results and their subsequent theoretical interpretation, continued experimental investigations into these systems are required to better understand the behaviour of HCIs in the cosmos.

\subsection{Measuring Decay-Rate Changes with TITAN}
To observe EC decay suppression under stellar conditions, a number of case-specific selections must be made to maximize the detection signal.  The top priority for the selection of suitable cases is based on the astrophysical importance for understanding decay-rate changes in hot environments.  Further consideration is required to determine which isotopes are best suited to the in-trap measurement technique, ie. short-lived (t$_{1/2}<1$~d) isotopes which are or can be produced at ISAC with high-rates and -purity, as well as those which have a reasonably high EC-branch.

For the measurements themselves, the ions are injected into the EBIT in bunches as SCIs from the TITAN RFQ at regular intervals which are typically on the order of $\sim50-100$~ms.  As the ions are injected into the EBIT, the inner electrode potential is lowered for first ion-bunch injection and subsequently raised to confine the first ion bunch.  Following this, the injected ion bunch(es) quickly reach $q>2^+$ and remain confined during subsequent injections due to the increased effective potential experienced by the newly formed HCIs.  Once the space-charge limit of the EBIT has been reached, the ions are trapped for up-to minutes, and the EC decays are observed by the characteristic X- and $\gamma$-rays resulting from the de-excitation processes.  The energies of the detected photons are case specific and can range from a few keV to 1~MeV.  Each detected photon is time stamped, and so the evolution of the number of photons as a function of time provides an excellent measurement for the half-life of the specific decay process being studied.

The primary difficulty in performing these studies inside the EBIT is that several charge states exist simultaneously in the trap, and must be disentangled during analysis.  To partially mitigate these effects, the ions are periodically ejected from the EBIT to an MCP downstream in the TITAN beamline where the number of different ions and charge states can be separated.  Furthermore, by performing these decay measurements at a variety of electron-beam energies, relative half-lives can be extracted based on the specific experimental conditions.

\section{Selective Blocking of Radioactive Decay Modes to Expose Second Order Processes}
\begin{figure}[t!]
\centering
\includegraphics[width=0.7\linewidth]{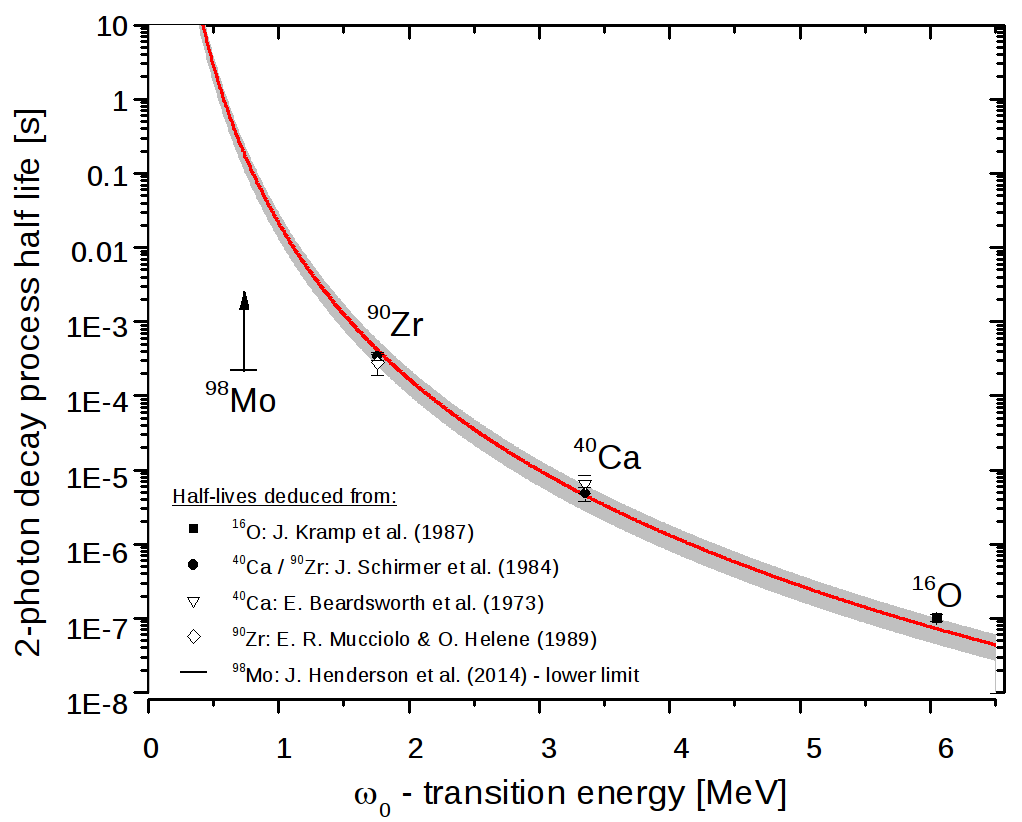}
\caption{Measured half-lives (or limits) for the 2$\gamma$-decay process deduced from total decay half-lives and 2$\gamma$ branching-ratios.  The red curve is a fit to the data with an $\omega_0^{7}$ dependence from Eqn.~\ref{eq:2gammawidth}.  The grey band indicates the region of 95\% confidence.}
\label{fig:2gammadecaydeduced}
\end{figure}

To first order, excited nuclear states release energy via single $\gamma$ emission with possible IC and internal pair creation (IPC) in competition.  Other modes of internal nuclear decay are rare and difficult to study as they involve the inclusion of higher-order processes in electroweak decay theory, and are thus significantly less likely.  Particularly of note is the nuclear two-photon (2$\gamma$) decay, a second order electroweak process which has only been observed in a few cases (eg.~\cite{Kra87,Wal15}).  The de-excitation through emission of two simultaneous photons is often represented as a virtual excitation of the initial state to a higher lying state followed by its de-excitation to the final state.  The width for this process can be approximated by,

\begin{equation}
\label{eq:2gammawidth}
 \Gamma_{2\gamma} = \frac{\omega_0^7}{105\pi} \left(   \alpha^2_{E1} +\chi_{M1}^2 + \frac{\omega_0^4}{4752} \, \alpha^2_{E2}    \right)
\end{equation}

\noindent where $\omega_0$ is the transition energy, $\alpha_{E1}$ and $\alpha_{E2}$ are the electric dipole and quadrupole transitions polarizabilities, respectively, and $\chi_{M1}$ is the magnetic dipole transition susceptibility.  These fundamental nuclear-structure observables are not easily accessible and usually require complex experiments to probe them \cite{Tam11,Kra87}.  Thus, the idea of obtaining them through the measurement of a 2$\gamma$ transition is very appealing~\cite{Hay90,Wal15}. If accessible through decay spectroscopy, the angular correlations between the two photons and the photon continuous energy spectrum greatly help on the determination of those quantities by fitting the full distribution~\cite{Kra87},

\begin{equation}
\label{eq:2gammadistr}
\begin{split}
 \frac{d^2 \Gamma_{2\gamma}}{d(\cos \theta) \, d \omega_1}= \frac{\omega_1^3 \, \omega_2^3}{\pi} \left[ \alpha^2_{E1} +\chi_{M1}^2 +
 \frac{\omega_1 \, \omega_2 \, \alpha_{E2}  \,\chi_{M1} }{6}   +  \frac{\omega_1^2 \, \omega_2^2 \, \alpha^2_{E2}}{144} 
 + ( 4 \, \alpha_{E1}  \,\chi_{M1}) \cos \theta  \right. \\
 +  \left(  \alpha^2_{E1} +\chi_{M1}^2  - \frac{\omega_1 \, \omega_2 \, \alpha_{E2}  \,\chi_{M1} }{2}   -  \frac{ 3 \, \omega_1^2 \, \omega_2^2  \, \alpha^2_{E2}}{144}    \right) \cos^2 \theta  \\
 - \frac{\omega_1 \, \omega_2 \, \alpha_{E1}  \, \alpha_{E2} }{3} \cos^3 \theta 
 \left.  + \frac{4 \, \omega_1^2 \, \omega_2^2 \, \alpha^2_{E2}}{144} \cos^4 \theta \right] 
\end{split}
\end{equation}

\noindent where $\omega_1$ and $\omega_2$ are the energies of the two emitted photons, which sum up to the transition energy $\omega_0$, and $\theta$ is the emission angle between them.  Due to the high level of competition in the decay modes, perhaps the best way to study this rare process is to avoid or suppress the other processes: one-photon emission, IPC and IC.  To date, studies on two-photon decay have focused on cases in which the one-photon emission is forbidden or heavily suppressed\footnote{The most recent result, however, reported the decay of an $11/2^-$ isomer of $^{137}$Ba to its $3/2^-$ ground state, and is the only {\it competitive} 2$\gamma$ emission ever measured~\cite{Wal15}.}.  This typically occurs in cases where both the ground and first-excited states are $J^\pi=0^+$, since the $E0$ photon emission is forbidden.  This is referred to as {\it non-competitive} 2$\gamma$ decay, and has only been observed in $^{16}$O~\cite{Kra87}, $^{40}$Ca~\cite{Sch84}, and $^{90}$Zr~\cite{Sch84,Muc89} (Fig.~\ref{fig:2gammadecaydeduced}).  An upper-limit on the $2\gamma$-decay branch in $^{98}$Mo is also reported in Ref.~\cite{Hen14}.  Although uncommon, several other isotopes share the same property of $0^+$ ground and first excited states (Table~\ref{tab:2gamma}).

\begin{table}[t!]
\small
\centering
\caption{A list of 2$\gamma$-decay candidates that have both $0^+$ ground and first excited states~\cite{NNDC}. The $\beta$-decay properties of potential parent nuclides used for production of these $0^+$ quasi-isomeric states using the method presented below are also shown.}
\label{tab:2gamma}
\begin{tabular}[h]{ccccc|ccc}
\hline\hline
$2\gamma$ Nucleus	&  $T_{1/2}^{gs}$ [s] 	&  $\omega_0$ [MeV]	&   $T_{1/2}^{0_2^+}$ [ns] 	&  $\Gamma_{2\gamma}/\Gamma_{total}$  & $\, \,$Parent Nucleus$\, \,$ &  $T_{1/2}$ [s] & BR to $0_2^+$ 	\\ \hline
$^{16}$O	& (stable)		& 6.049		& 0.067(5)		& $6.6(5)\cdot 10^{-4}$	& $^{16}$N	& 7.13	& $1\cdot 10^{-4}$	\\
$^{40}$Ca	& (stable)		& 3.353		& 2.16(6)		& $4.5(10)\cdot 10^{-4}$	& $^{40}$Sc	& 0.182	& (unknown)	\\
$^{90}$Zr	& (stable)		& 1.761		& 61.3(25)		& $1.8(1)\cdot 10^{-4}$	& $^{90}$Nb	& $5.3\cdot 10^{4}$	& $ 5\cdot 10^{-5}$	\\
		& 			& 		& 			& 				& $^{90}$Y	& $2.3\cdot 10^{5}$	& $1.2\cdot 10^{-4}$	\\ \hline
$^{68}$Ni	& 29(2)			& 1.770		& 270(5)		& -	& $^{68}$Co	& 0.20	& (unknown)	\\
$^{72}$Ge	& (stable)		& 0.691		& 444.2(8)		& -	& $^{72}$Ga	& $5.0\cdot 10^{4}$	& $ 5.0\cdot 10^{-3}$	\\
$^{80}$Ge	& 29.5(4)		& 0.639		& -		& -	& $^{80}$Ga	& 1.9	& (unknown)	\\
$^{96}$Zr	& (stable)		& 1.581		& 38.0(7)		& -	& $^{96}$Y	& 5.34	& $ 1.4\cdot 10^{-2}$	\\
$^{98}$Zr	& 30.7(4)		& 0.854		& 64(7)			& -	& $^{98}$Y	& 0.548	& $ 1.5\cdot 10^{-1}$	\\
$^{98}$Mo	& (stable)		& 0.735		& 21.8(9)		& -	& $^{98}$Nb	& 2.86	& $ 2.6\cdot 10^{-1}$	\\ \hline\hline
\end{tabular} 
\end{table}


As with the EC-decay studies presented in the previous section, further case-selection criteria must also be applied before these measurements are possible.  For instance, the only way to avoid IPC as a dominant mode of decay is to explore cases where it is energetically forbidden (ie. $\omega_0 < 1.022$ MeV).  However, even avoiding cases in which one-photon emission and IPC are present, decay from these excited $0^+$ states still occurs via IC with a much higher probability than 2$\gamma$ decay.  A novel method to eliminate the highly competing $0^+\rightarrow0^+$ IC has therefore been devised to elucidate the rare $2\gamma$ process.  As described in the previous section for EC, the complete suppression of this mode is possible if the ion is fully stripped, thus allowing for 2$\gamma$ decay to be probed with no competition.  The recently commissioned decay spectroscopy setup with the TITAN EBIT provides a unique environment for generating these conditions and observing such decays~\cite{Lea15}.  The technique for performing these measurements is described below.

\subsection{Experimental Method for $2\gamma$ Decay in the TITAN EBIT}
The excited $0^+_2$ state of the isotope of interest must be populated using a suitable mechanism such that the ion remains in its bare-ion configuration.  In the TITAN EBIT, this is done via the $\beta$ decay of a parent ion (see Table~\ref{tab:2gamma}) delivered by TRIUMF-ISAC.  The general principle is that the parent ion is injected into the TITAN EBIT, removed of all bound electrons, and subsequently decays to the $0^+_2$ state of the daughter nucleus which remains stable unless it undergoes $2\gamma$ decay.  Since the expected 2$\gamma$-decay half lives are much lower than the maximum trapping time in the EBIT, the highly charged ion-cloud - now rich in $0^+_2$ {\it quasi-isomers} that will completely decay inside EBIT exclusively through 2$\gamma$ - can be studied using the photon detector setup.  The clean in-trap environment allows the measurement of spectra with very low background; and seven HPGe detectors placed at view ports could probe angular correlations at 45$^\circ$, 90$^\circ$, 135$^\circ$, and 180$^\circ$, with a total two-photon coincidence detection efficiency of about $2\times10^{-5}$. For a typical one-week RIB experiment, this level of sensitivity is sufficient to measure energy and angular distributions to determine the nature of the transition, and the ground state polarizabilities and susceptibilities.

One clearly identified limitation of the current system, however, is the low geometric coverage of the photon detectors due to the limited optical access to the trap center ($\sim2\%$ of the total $4\pi$ solid angle~\cite{Lea15}).  This limits the number of angles that can be used for angular correlations, and thus limits the statistics that can be used for extracting the physically relevant quantities from Eqn.~\ref{eq:2gammadistr}.  Additionally, since the HPGe photon detectors are single crystals and not clover or cluster-style detectors with several crystals packed together, information on the multipolarity of the $\gamma$ rays are currently not accessible through the traditional method of Compton scattering reconstruction.  As a part of the ongoing upgrades outlined in the following section, new state-of-the-art photon detector solutions for extracting the quantities of interest are being explored.

\section{Conclusions and Future Work}
In summary, the recently commissioned TITAN EBIT decay spectroscopy setup provides a unique environment for probing electroweak decay properties using HCIs.  The current experimental program is based on studying EC in hot astrophysical environments, as well as rare second-order decay processes to probe fundamental nuclear structure observables.  The TITAN facility is currently undergoing multiple improvements to the experimental system~\cite{Lea15c}, including an upgrade of the EBIT electron-gun to 60~keV at 2~A to increase our ability to fully ionize the radioactive samples of interest to our scientific program.  The infrastructure for this upgrade at TITAN has been completed, and the new gun is currently being constructed at the Max-Planck Institute for Nuclear Physics in Heidelberg, and will be delivered in 2017.  Additionally, the installation of an MR-ToF into the TITAN system will be performed in early 2017 which will improve the sample purity inside the EBIT by performing isobaric separation at the level of roughly 1 part in $\sim10^4$ for the cases and ion numbers of interest to this program\footnote{The MR-ToF device has already demonstrated a mass resolving power of $2\times10^5$ for low ion numbers~\cite{Jes15}.}.  An upgrade of our photon detectors to include higher efficiency through the use of clover-style HPGe detectors is also planned to further increase the sensitivity for low-rate experiments.  To address the issue of multiple charge-state populations within the EBIT during the decay measurements, plans for a dedicated decay spectroscopy trap external to the EBIT are currently underway.  An external trap would allow for specific isotope and charge-state selection following the extraction of ions from the EBIT.

\vspace{6pt} 

\acknowledgments{The authors would like to acknowledge the efforts of the TITAN collaboration towards the development of this program.  This work is supported in part by the National Science and Engineering Research Council of Canada (NSERC), the Deutsche Forschungsgemeinschaft (DFG) under grant FR 601/3-1, and Brazil's Conselho Nacional de Desenvolvimento Cient\'ifico e Tecnol\'ogico (CNPq).  TRIUMF receives federal funding via a contribution agreement with the National Research Council of Canada (NRC).}

\authorcontributions{The scientific direction for the HCI decay program presented here has been developed and conceived of by K.G.~Leach, I.~Dillmann, R.~Klawitter, and E.~Leistenschneider.  The technical development of the EBIT for HCI decay spectroscopy has been primarily performed by K.G.~Leach, A.~Lennarz, and R.~Klawitter.  The technical development of the co-existing "TITAN-ec" program $-$ thus facilitating the current work on HCIs $-$ was performed by A.~Lennarz, T.~Brunner, K.G.~Leach, D.~Frekers, and C.~Andreiou.  A.A.~Kwiatkowski oversees the operation of the TITAN ion trap system.  J.~Dilling is the principal investigator of TITAN and the head of the collaboration.}

\conflictofinterests{The authors declare no conflict of interest.} 

\bibliographystyle{mdpi}


\bibliography{references}

\end{document}